\documentclass[aps,reprint,amsmath,amssymb]{revtex4-1}

\usepackage[dvips]{graphicx}
\usepackage{psfrag}
\usepackage{rotating}
\usepackage{natbib}
\usepackage{dcolumn}
\usepackage{bm}
\usepackage{MR}

\begin{document}

\title{
On the excitation of inertial modes in an experimental spherical
Couette flow}

\author{Michel Rieutord}

\affiliation{
Universit\'e de Toulouse; UPS-OMP; IRAP; Toulouse, France\\ and
CNRS; IRAP; 14 avenue E. Belin, 31400 Toulouse, France
}
\email{Michel.Rieutord@irap.omp.eu}

\author{Santiago Andr\'es Triana}
\affiliation{Department of Physics and Institute for Research in Electronics and Applied Physics, University of Maryland, College Park, MD 20742, USA}
 \email{triana@umd.edu}

\author{Daniel S. Zimmerman}
\affiliation{Department of Physics and Institute for Research in Electronics and Applied Physics, University of Maryland, College Park, MD 20742, USA}
 \email{danzimmerman@gmail.com}

\author{Daniel P. Lathrop}
\affiliation{Department of Physics, Department of Geology, Institute for Physical Science and Technology, and Institute for Research in Electronics and Applied Physics, University of Maryland, College Park, MD 20742, USA}
 \email{lathrop@umd.edu}

\date{\today}

\begin{abstract}
Spherical Couette flow (flow between concentric rotating spheres) is
one of flows under consideration for the laboratory magnetic dynamos.
Recent experiments have shown that such flows may excite Coriolis restored
inertial modes. The present work aims to better understand the properties
of the observed modes and the nature of their excitation. Using numerical
solutions describing forced inertial modes of a uniformly rotating fluid
inside a spherical shell, we first identify the observed oscillations
of the Couette flow with non-axisymmetric, retrograde, equatorially
anti-symmetric inertial modes, confirming first attempts using a
full sphere model. Although the model has no differential rotation,
identification is possible because a large fraction of the fluid in a
spherical Couette flow rotates rigidly. From the observed sequence of the
excited modes appearing when the inner sphere is slowed down by step, we
identify a critical Rossby number associated with a given mode and below
which it is excited. The matching between this critical number and the
one derived from the phase velocity of the numerically computed modes
shows that these modes are excited by an instability likely driven by
the critical layer that develops in the shear layer staying along the
tangent cylinder of the inner sphere.

\end{abstract}

\maketitle

\section{Introduction}

Large-scale flows in stars or planets in many circumstances take place
in a spherical shell. Most astrophysical fluid flows are also under the
dominating influence of a background rotation. This rotation leads to
the presence of inertial oscillations for which the restoring mechanism
{comes from the} Coriolis acceleration.

The properties of these modes of oscillation are not fully
understood, in part because they obey a hyperbolic equation in the space
variables and therefore do not easily comply with boundary conditions.
The solutions of this equation, known as the Poincar\'e equation, have
been studied in some details in the recent years, using numerical and
analytical tools \cite[e.g.][]{ML95,RV97,RGV01,RVG02,O05,RV10}. It has
been shown that most of the eigenmodes of a rotating spherical fluid
layer require viscosity to exist.  Indeed, viscosity is necessary to
regularize the singularities formed by the focussing of characteristics
by the boundaries. Viscosity transforms these singularities into
shear layers whose thickness scales with some fractional power of it
(exponents 1/4 or 1/3 are the most common). In addition, very recent works
\cite[][]{GL09,RV10} showed that the critical latitude singularity
where the characteristics of the hyperbolic equation are tangent to
the inner core boundary, plays a crucial role in periodically forced
flows. The reason for that is not clear presently.

However, all the aforementioned previous works are theoretical studies
considering idealized situations, and therefore should be compared
to experimental studies.  Observations of inertial modes are not
extensive, either in Nature or in the laboratory. A landmark in the
experimental studies is the work of Aldridge\cite{AlToo69,Ald72}. More
recently, attractors of characteristics triggering oscillatory shear
layers have been investigated experimentally for the understanding
of ocean dynamics. Such experiments were conducted both on stably
stratified fluids \cite[][]{MBSL97,HBDM08,HTM10} and rotating fluids
\cite[][]{Maas01,MM04,Maas05} since internal modes (gravity waves)
and inertial modes share many of the same mathematical properties.
Other experiments have demonstrated the excitation of inertial waves in
a fluid inside a precessing spheroidal cavity \cite[][]{Noir2001a}.

\begin{figure}
\centerline{
\includegraphics[width=\linewidth,clip=true]{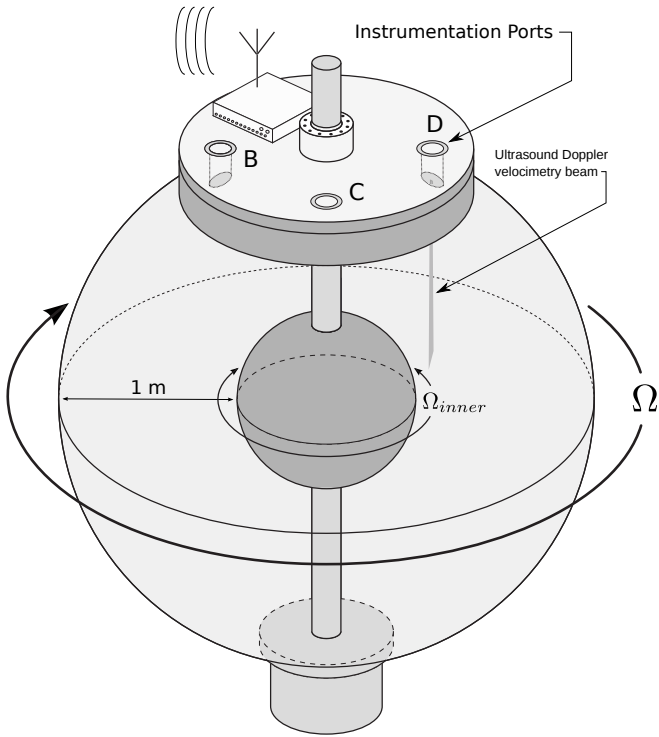}}
\caption[]{Schematic of the 3-meter spherical-Couette apparatus. Inner
and outer spheres rotate independently, driven by two 250 kW motors (not
shown). There are three pressure sensors (azimuthally $90^\circ$ apart) on the top
lid ports, and an ultrasound velocimetry transducer measuring vertical
components of fluid velocities (ultrasound beam depicted coming from
port D).}
\label{expsetup}
\end{figure}

\begin{figure}
\centerline{
\includegraphics[width=\linewidth,clip=true]{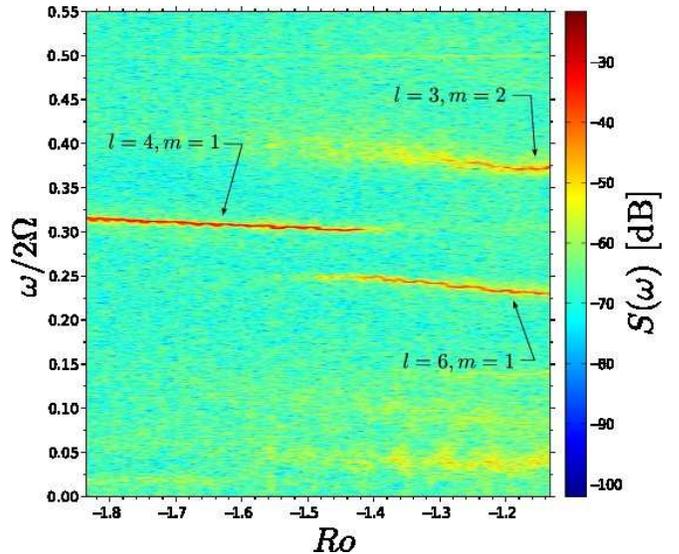}}
\caption[]{(Color online) Spectrogram from pressure measurements as the inner
sphere rotates with different speeds in counter-rotation (inner sphere
rotating in the opposite direction as
the outer sphere). Each vertical line in the spectrogram is the power
spectral density of the pressure using $\rho \RO \Omega^2 L^2$ as the
unit pressure. Outer sphere rotation rate is 1.5 Hz corresponding to
$E=2.5\times10^{-8}$. Modes indicated correspond to full sphere modes
characterized by $(l,m,\omega/2\Omega)$.}
\label{spectrogram1}
\end{figure}

\begin{figure}
\centerline{
\includegraphics[width=\linewidth,clip=true]{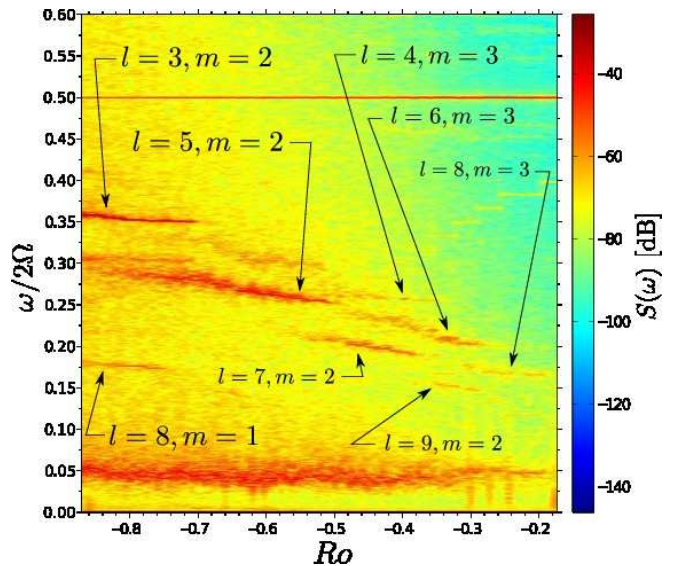}}
\caption[]{(Color online) Same as Fig.~\ref{spectrogram1} but for a $\RO$ range
corresponding to co-rotation (inner sphere rotating in the same direction
as the outer). The band near $\omega/2\Omega =0.05$, which has an azimuthal number
$m=1$, is possibly not a single inertial mode.
It's signature in the experiment reported by
Kelley et al. \cite[][]{KTZL10} is weak, perhaps a consequence of its
quasi-geostrophic character (see discussion at end of Sect. IIIB).}

\label{spectrogram2}
\end{figure}

All these experiments have shown that inertial modes are robust
features of rotating fluid flows. In a very recent experiment aimed at
studying a fluid dynamo, inertial modes were detected through their
coupling with an imposed magnetic field, in a spherical Couette flow
\cite[][]{KTZL10}. This flow can indeed produce magnetic fields at
sufficiently high magnetic Reynolds numbers \cite[][]{GC10}. In that
experiment, the fluid was contained in a spherical shell with inner and
outer radii equal to 10~cm and 30~cm respectively. The Ekman number,
the non dimensional measure of viscosity (see below), was approximately
10$^{-7}$.  In astrophysical or geophysical {(Earth's core)} situations
this number is rather {less} than 10$^{-12}$. In fact, as was shown
by recent numerical work \cite[e.g.][]{RGV01,RV10}, the asymptotic
regime describing vanishingly small viscosities usually appears at Ekman
numbers below 10$^{-8}$. A new experiment geometrically similar to
the Earth's core, using a sphere with an outer radius of 1.46m offers a
unique opportunity to observe some near singular inertial modes close to their
asymptotic regime since now Ekman numbers can be as low as $2.5\times
10^{-8}$. Indeed, first results on this experiment using a  precessionally
forced flow \cite[][]{TZL12} provided a clear evidence of detached
shear layers spawned by the critical latitude singularities.

\begin{figure*}
\centerline{
\includegraphics[width=0.50\linewidth,clip=true]{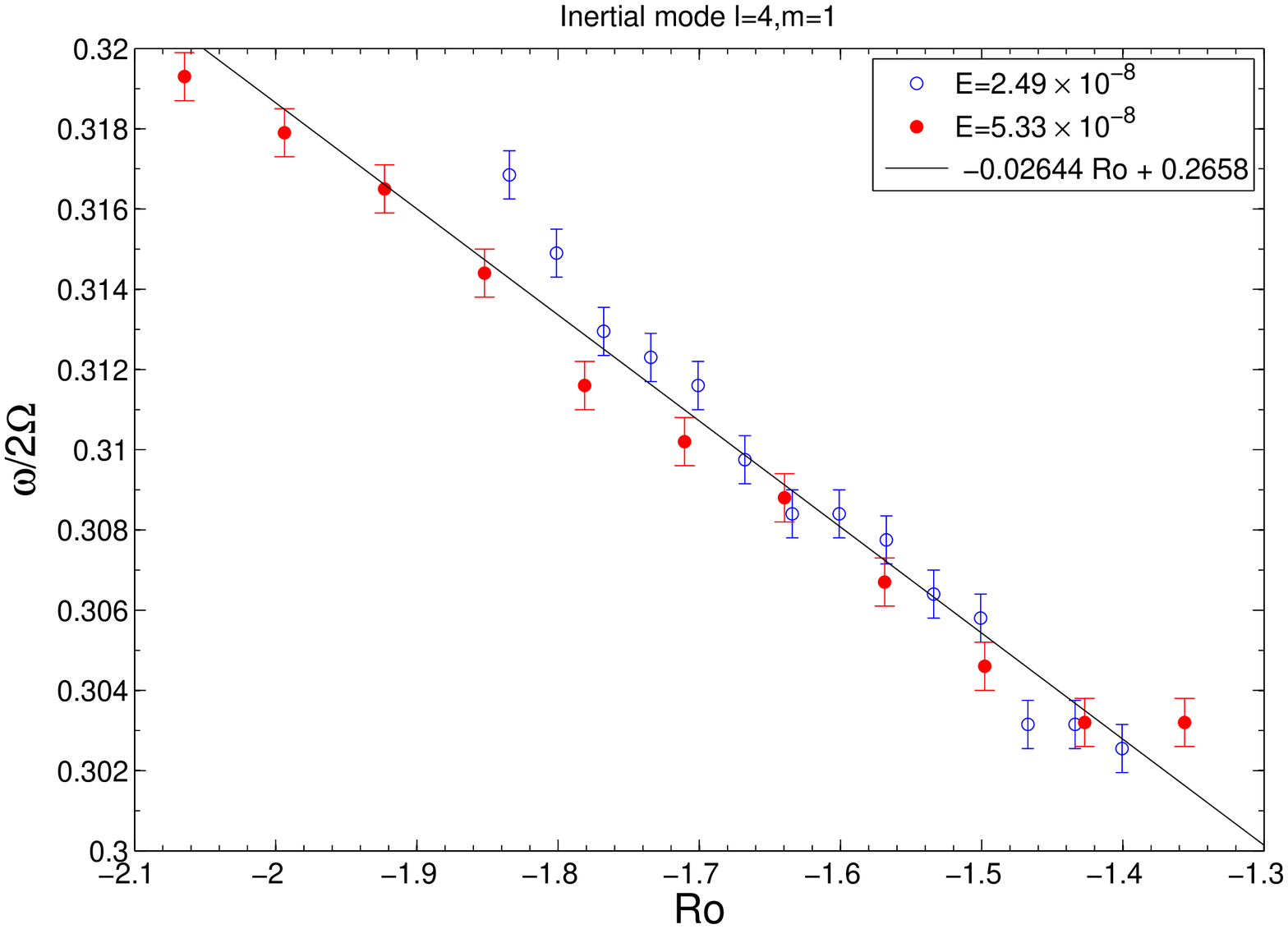}
\includegraphics[width=0.50\linewidth,clip=true]{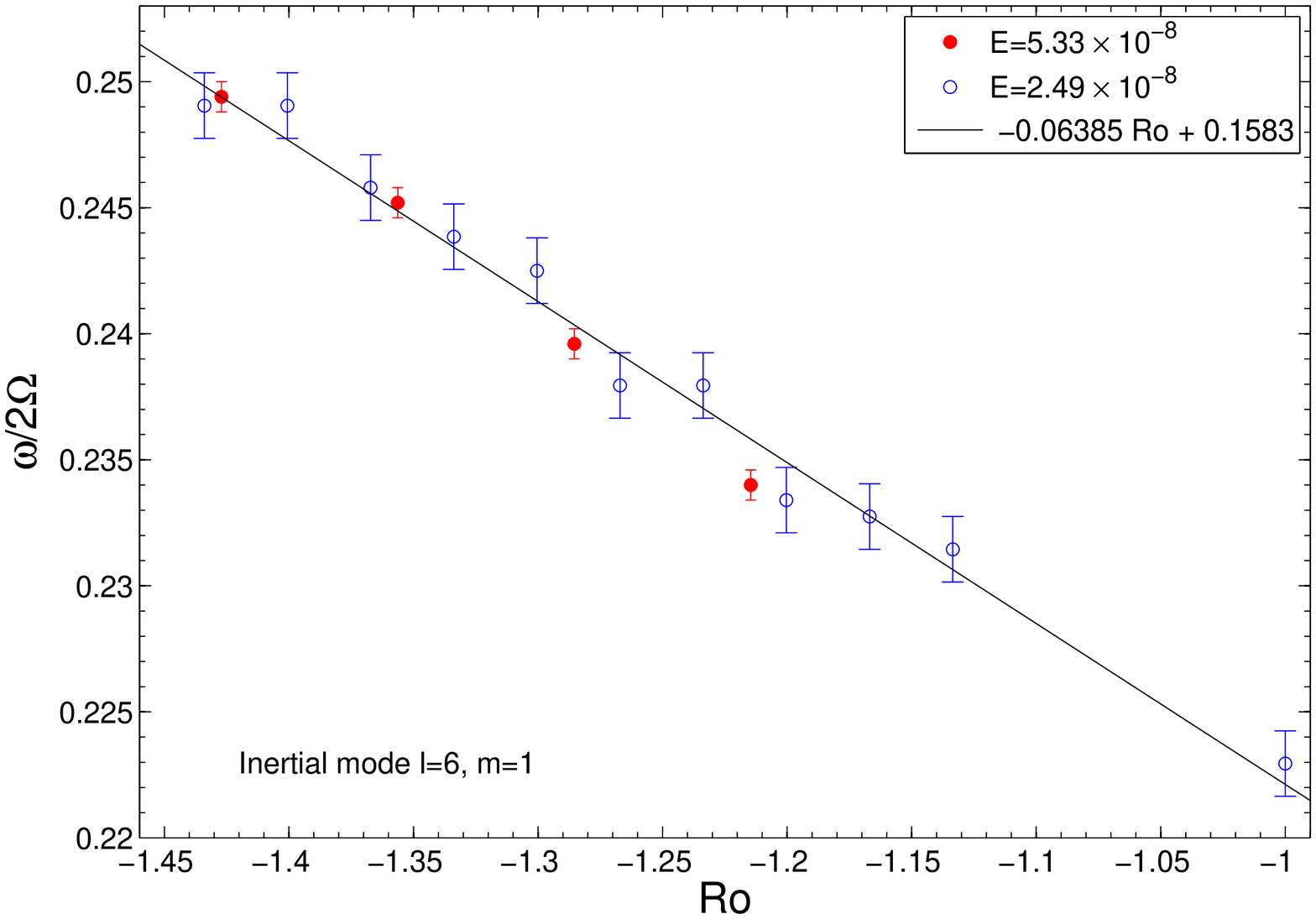}
}
\caption[]{(Color online) Rossby number dependence of the frequency of the two most prominent
m=1-modes. The insert gives the equation of the linear best fit.}
\label{omega_ro1}
\end{figure*}

\begin{figure*}
\centerline{
\includegraphics[width=0.33\linewidth,clip=true]{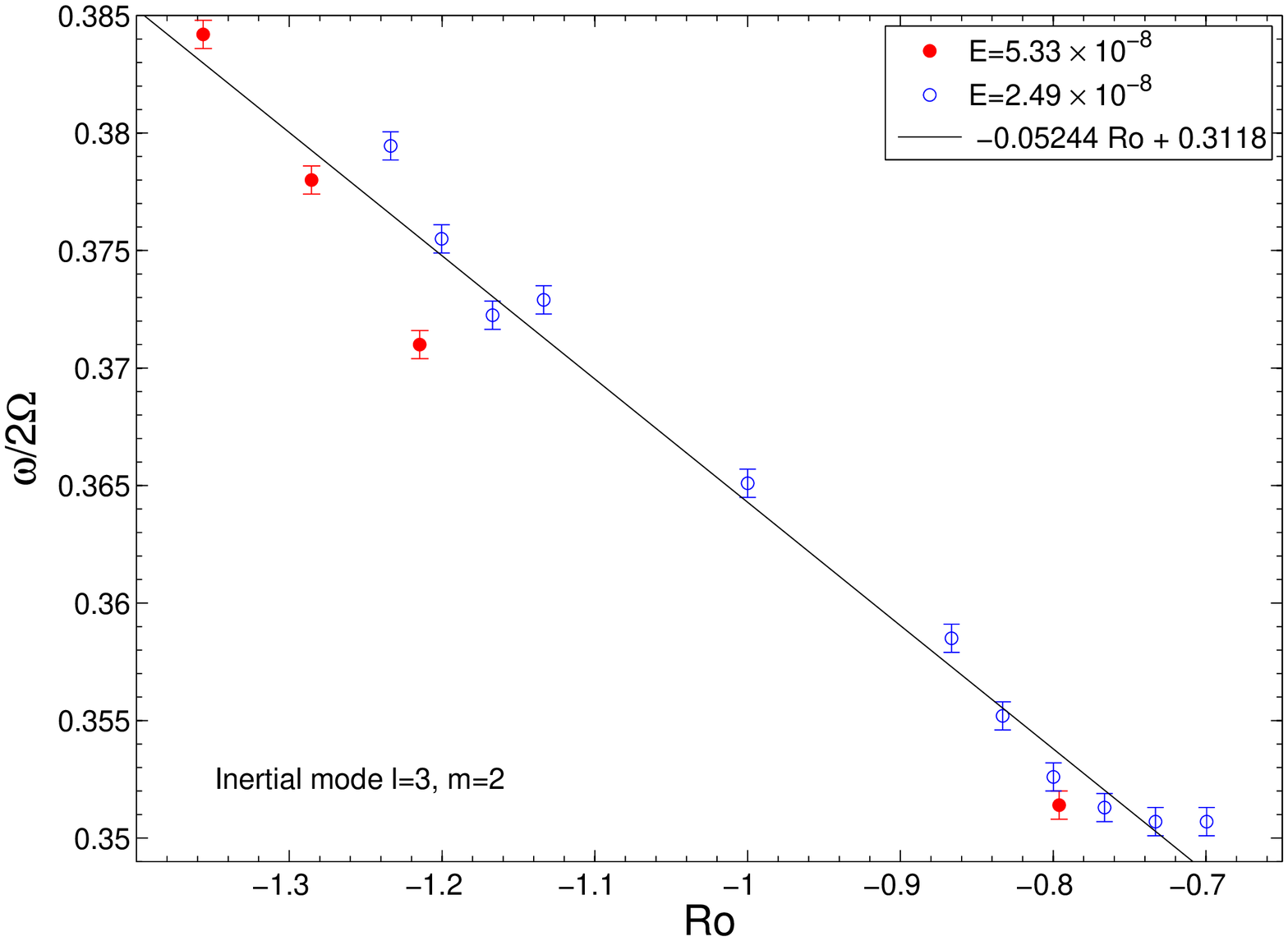}
\includegraphics[width=0.33\linewidth,clip=true]{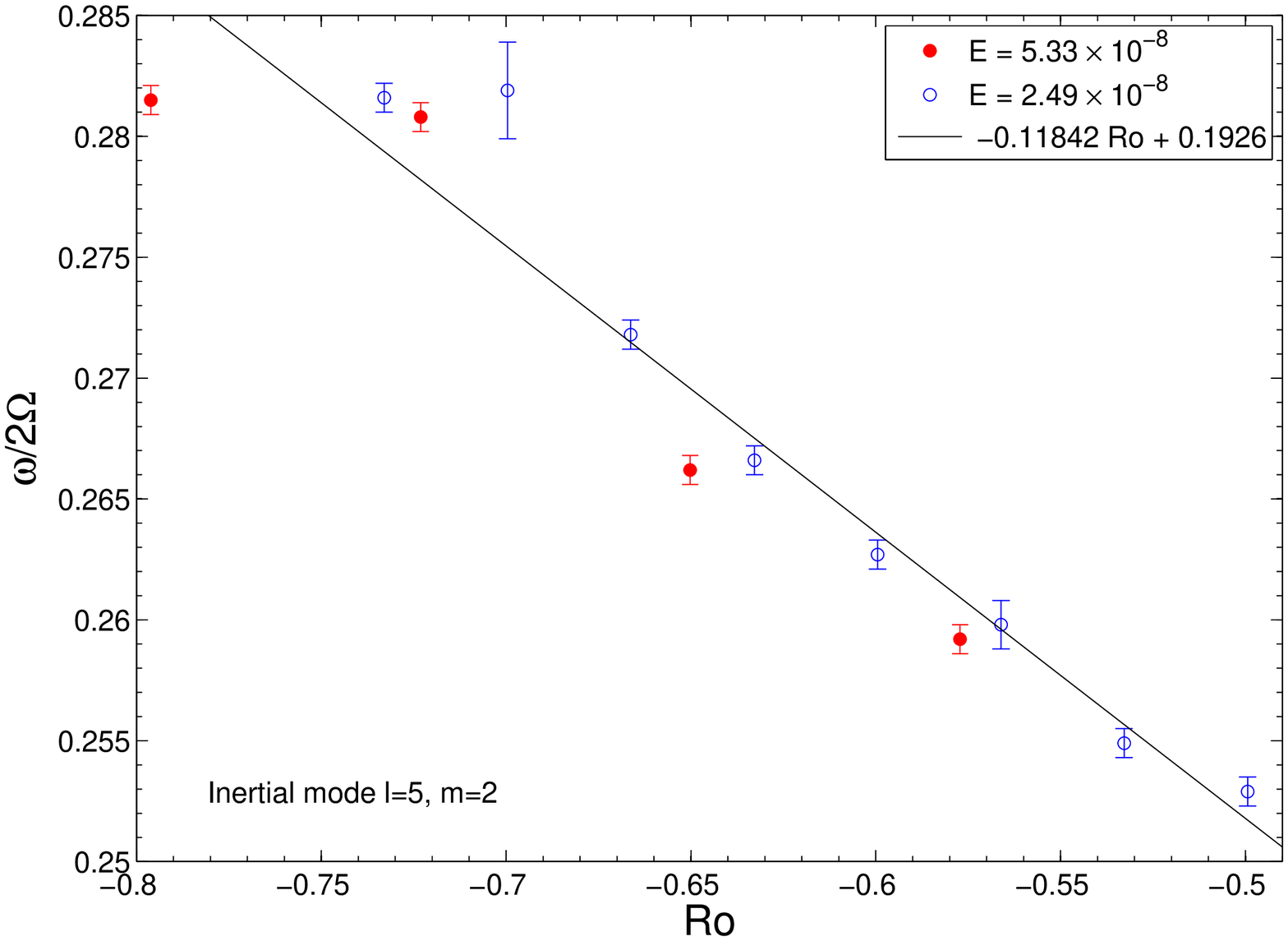}
\includegraphics[width=0.33\linewidth,clip=true]{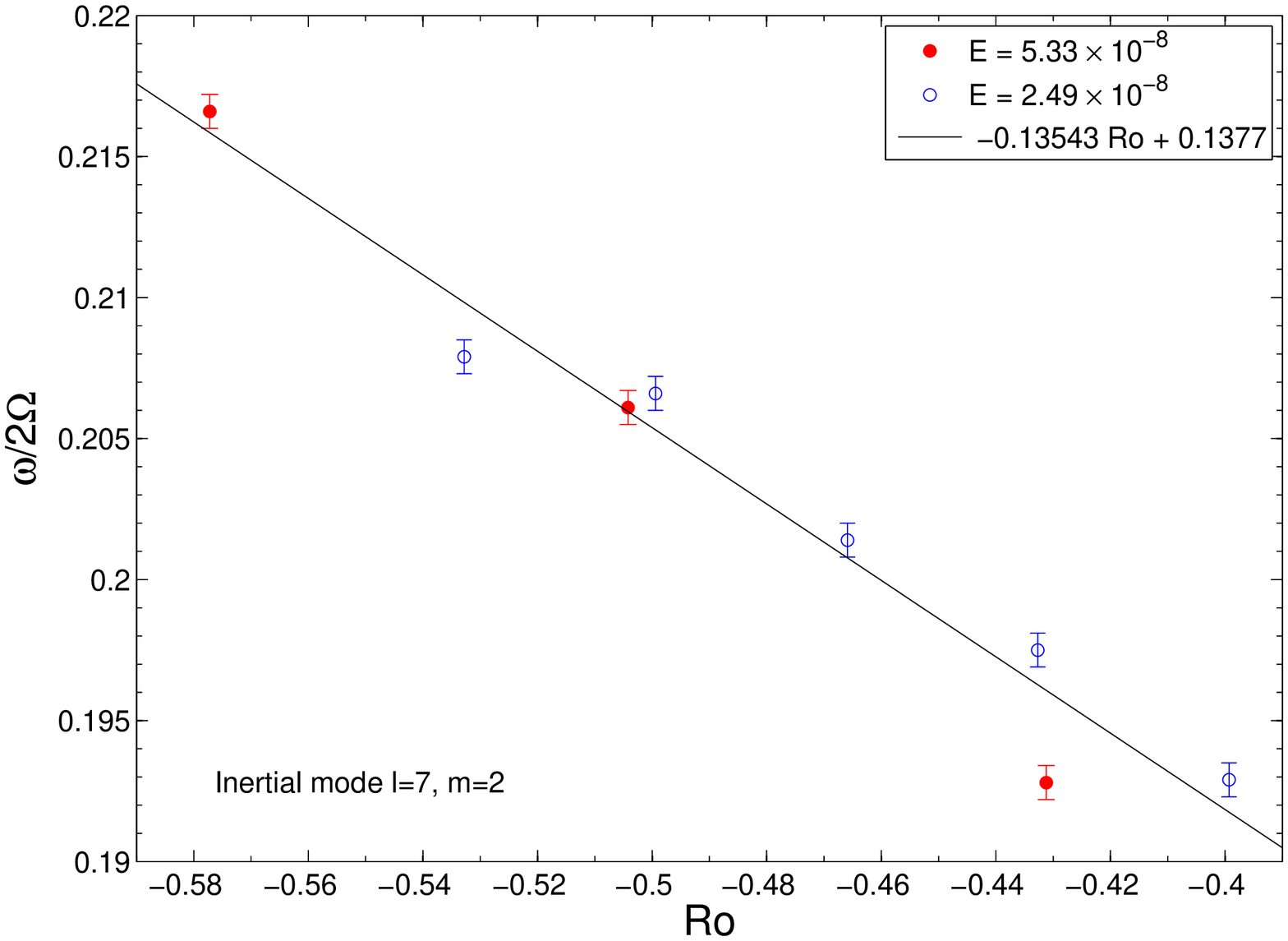}
}
\caption[]{(Color online) Same as Fig.~\ref{omega_ro1} but for the three most prominent m=2-modes.}
\label{omega_ro2}
\end{figure*}

In this paper we further consider the results of this experiment. Thanks
to a simple model based on numerical solutions of forced inertial modes
in a spherical shell, we find a scenario that explains the excitation of
inertial modes in a nonlinear spherical Couette flow.  For this,
we first describe the experimental set-up and the observational facts
concerning the observed inertial modes (sect.2). We then compute the
response of an incompressible fluid inside a rigidly rotating spherical
shell when some periodic forcing is applied (sect.3). We use this simple
model to identify the resonance peaks and to interpret their full width at
half-maximum (FWHM). We then discuss a scenario that explains most of
the experimental facts (sect. 4). Conclusions end the paper.

\section{The experimental set-up and observations}\label{exp}

The 3-meter Geodynamo Experiment built at the University of Maryland
consists of a rotating, stainless-steel spherical vessel with 1.46 m
of radius, and an independently rotating 1.02 m diameter inner sphere,
located at the center of the outer spherical vessel (see a schematic view
in Fig.~\ref{expsetup}). This results in
an aspect ratio $\eta\simeq0.348$. There is a 16.8 cm diameter shaft
supporting the inner sphere that extends along the axis of the outer sphere and
is connected to a motor. The outer sphere is driven by a
motor via a geared ring attached to the top lid. The space between the
spheres is filled here with water at room temperature (kinematic viscosity
$\nu=1.004\times 10^{-6}$ $m^2/s$). The Rossby number is defined as
$\RO=(\Omega_{inner}/\Omega)-1$ where $\Omega_{inner}$ is the angular
speed of the inner sphere and $\Omega$ is the angular speed of the
outer sphere. The outer sphere can spin at a maximum of 4 Hz
and 0.05 Hz minimum, although for the measurements presented here the
maximum rotation rate corresponds to 1.5 Hz. Defining the Ekman number as
$E=\nu/2\Omega R^2$, where $R=1.46$m, the experimentally accessible range
is $2.5\times 10^{-8}<E<7.5\times10^{-7}$. The inner sphere can reach
a rotation rate up to 20 Hz with a minimum of 0.15 Hz which translates
into a Rossby number range such that $0.03\lesssim|\RO+1|\lesssim400$.

\begin{figure}
\centerline{
\includegraphics[width=\linewidth,clip=true]{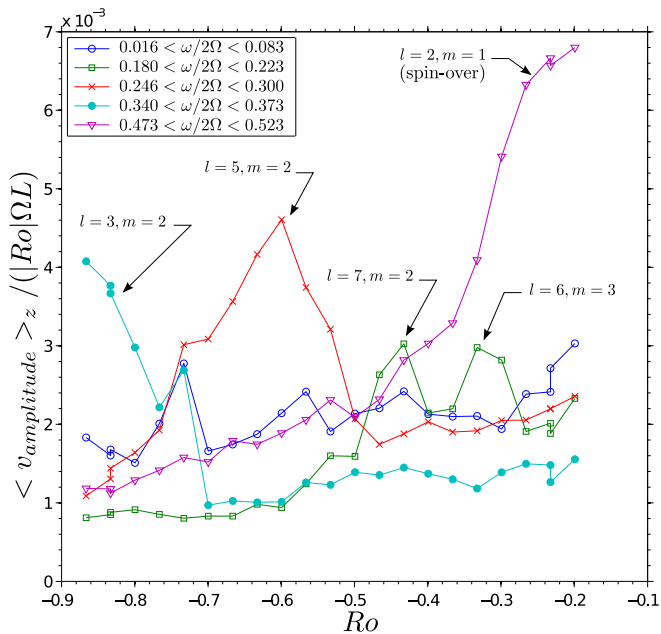}}
\caption[]{(Color online) Velocity amplitude averaged over a thin layer, perpendicular
to the rotation axis, located around $z=0.846$ as a function of the
Rossby number. Each curve corresponds to the averaged velocity amplitude
restricted to the frequency range indicated in the legend. Labels
correspond to identified inertial modes.}
\label{obs_velo}
\end{figure}

During the experimental run corresponding to the measurements
shown below, the outer sphere had a fixed rotation rate of 1.5 Hz
($E=2.5\times10^{-8}$) while the inner sphere angular speed was
varied. We started with the inner sphere counter-rotating (near
$\RO=-2$) and gradually reduced its angular speed in steps of 0.05 Hz
until its minimum speed was reached ($\RO \simeq -1.1$). At that point
the inner sphere rotation direction was changed and its speed was
increased from its minimum ($\RO \simeq -0.9$) until reaching almost
the same speed as the outer sphere ($\RO \simeq -0.1$). Each speed
step was maintained for about 15 minutes. Fig.~\ref{spectrogram1}
shows a power spectral density $S(\omega)$ spectrogram computed from
the pressure on one of the instrumentation ports, measured during
counter-rotation. Fig.~\ref{spectrogram2} shows a spectrogram of the
pressure measured during co-rotation.

The flow is monitored by pressure probes and an ultrasound velocimetry
transducer (see Fig.~\ref{expsetup}), all fixed to the outer
bounding sphere.  The pressure spectrograms show the signature of inertial
modes excited in a sequence as $\RO$ is varied. The azimuthal wave
numbers $m$ could be determined up to $m=4$ by comparing the relative
phases of the pressure signals from different ports.  These modes
match very closely the frequencies and azimuthal wave numbers of the
inertial modes excited in the hydromagnetic experiment performed by
Kelley et al. \cite{Kelley2007} using a 60-cm diameter sphere, which is
geometrically similar, but smaller than the 3-meter experiment.  We use
the spatial pattern data from the 60-cm experiment in order to identify
the modes in the 3-meter experiment.  Those modes, as evidences by lines
in the spectrogram, correspond approximately to full sphere inertial modes
\cite{ZELB01,Green69}, which can be characterized by a pair of indices
$(l,m)$ and a dimensionless frequency $\hat{\omega}=\omega/2\Omega$.
We also noticed {\em the retrograde propagation of the modes,} thanks to the
pressure probes distributed in longitude over the outer shell. This was
also the case for the previous (smaller) experiment reported by Kelley
et al. \cite[][]{KTZL10}.

The spectrograms show that the frequency of excited modes varies with
the Rossby number. In Fig.~\ref{omega_ro1} and \ref{omega_ro2} we plot
these variations for five modes and two values of the Ekman number. The
variations are approximately linear, showing an increasing frequency with
a decreasing (negative) Rossby number or, equivalently, with an increasing
differential rotation. A change of the Ekman number induces only a mild
change of the frequencies. We may however notice that the first of the
m=1-modes (Fig.~\ref{omega_ro1}) shows a steeper dependence with $\RO$
for the lowest value of the Ekman number.

The pressure spectrograms are complemented by velocity measurements
displayed in Fig.~\ref{obs_velo}. Those show the vertically averaged
velocity amplitude over a fluid layer located at $0.8325 < z/R < 0.8599$,
as a function of the Rossby number ($z$ is the coordinate along the
rotation axis, with $z=0$ corresponding to the equatorial plane). The
various curves show that some modes, already identified in the pressure
spectrograms, dominate the dynamics of the sampled layer for specific
ranges of the Rossby number. Especially, we observe that the spin-over
mode ($m=1$ and $\omega= -\Omega$) is always present and dominates the
oscillations for $\RO \supapp -0.4$.

\section{Numerical model}

\begin{figure}
\centerline{
\includegraphics[width=\linewidth,clip=true]{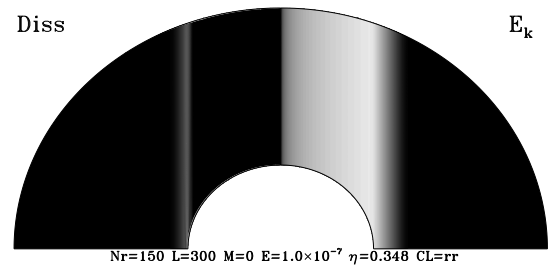}}
\caption[]{Numerically computed spherical Couette flow at very low
Rossby number $|\RO|\ll1$. This solution has been obtained by solving
steady, linear, axisymmetric equations of the spherical Couette flow using
a spectral method with spherical harmonics (horizontally) and Chebyshev
polynomials (radially). Left: the viscous dissipation emphasizing the
Stewartson shear layer. Right: the kinetic energy of the flow in the
reference frame of the outer shell.}
\label{couette_flow}
\end{figure}

The foregoing description of the experiment shows that despite a steady
forcing, the generated flows are unsteady \cite[][]{KTZL10}. This means
that the spherical Couette flow is unstable to time-dependence and the
excitation of inertial modes.

From the observed features of spherical Couette flow, which is shown in
Fig.~\ref{couette_flow} in the limit $|\RO|\ll1$, we expect that most
of the time-dependent forcing occurs in the fluid lying near or inside
the tangent cylinder. This cylinder is clearly marked by the vertical
Stewartson shear layer that exhibits large viscous dissipation. Most
of the shear is localized inside the Stewartson layer. This
is also approximately the case at finite Rossby numbers as shown by
Matsui et al. \cite{matsui_etal11}. However, the volume inside the
tangent cylinder is a rather small fraction of the total volume of
the fluid. This fraction reads

\[ 1-\frac{(1-\eta^2)^{3/2}}{1-\eta^3} \simeq 3\eta^2/2-\eta^3 \quad
{\rm if} \quad \eta\ll1\]
In the actual experiment where $\eta=0.348$ this fraction is
0.14. Thus, 86\% of the fluid lies outside the tangent cylinder and
rotates near solid body with the angular velocity of the outer shell
\cite[e.g.][]{stewar66}. This remark suggests that oscillations of a
rigidly rotating spherical fluid shell may not be very far from those
observed in the aforementioned experiment and that, as a first step,
differential rotation can be ignored.

\subsection{Mathematical formulation}

Assuming that the inertial modes are of relatively small amplitude,
they solve the linearised equations governing a viscous
rotating fluid with constant density.  We force periodic perturbations
in the numerical simulations to drive the modes.  With a length scale of
the outer radius of the shell $R$, the time scale as $(2\Omega)^{-1}$, the
equations of the non-dimensional pressure ($p$) and velocity perturbation
($\vu$) may be written:

\greq
i\hat{\omega}\vu + \ez\times\vu  = -\na p + E\Delta\vu \\
\\
\Div\vu = 0
\egreqn{eqmotion}
where $\hat{\omega}=\omega/2\Omega$ is a dimensionless frequency,
$E=\nu/2\Omega R^2$ is the Ekman number and $\nu$ is the kinematic
viscosity.

We complete these equations with no-slip boundary conditions which assists in
forcing the flow. Since the real forcing is not known, we drive oscillations
by setting a toroidal motion on one of the boundaries. Namely, we assume
that either on the inner or outer boundary

\beq \vu(\theta,\varphi) = \lp\dsphi{\YMM}\etheta-\dtheta{\YMM}\ephi\rp
e^{i\hat{\omega}\tau} \eeqn{bc1}
and

\beq \vu = 0 \eeqn{bc2}
on the other boundary. We note that a forcing by boundaries has  also been
used to force axisymmetric inertial modes \cite[][]{AlToo69,R91,tilg99b}.

\begin{figure}
\centerline{
\includegraphics[width=\linewidth,clip=true]{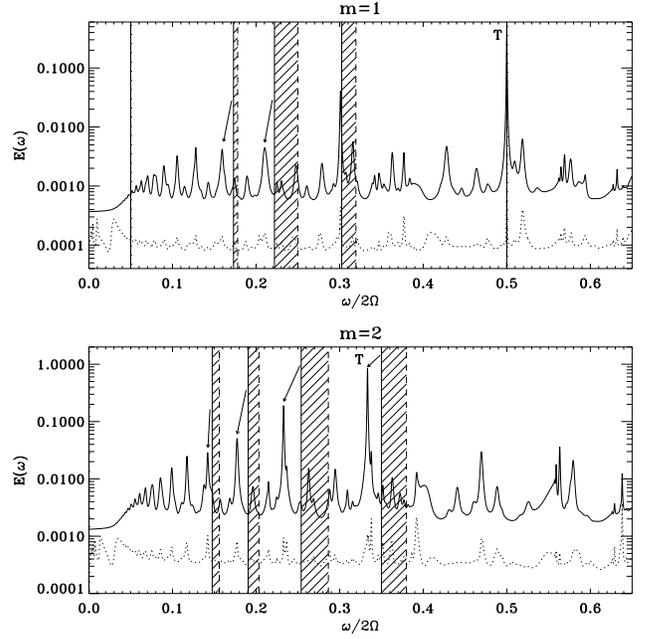}}
\caption[]{Resonance curves showing the kinetic energy (in arbitrary
units) of the model oscillating flow as a function of the scaled frequency
of the forcing. The solid line is for the forcing on the outer boundary,
the dotted line for a forcing on the inner shell. The hatched bands of
frequencies show the range of frequencies of the observed modes, when the
Rossby number is varied. The lower bound of the bands, materialized by
a solid vertical line, corresponds to the lowest $|\RO|$ value. The ``T"
marks the toroidal mode resonance. Arrows show the identifications
we suggest between observed oscillations frequencies and resonances in
the model. Top figure is for the $m=1$-forcing,
while the bottom figure is for an $m=2$-forcing. The Ekman number is
set to $E=2.5\times 10^{-8}$.}

\label{couette_flow1}
\end{figure}

\begin{figure}
\centerline{
\includegraphics[width=\linewidth,clip=true]{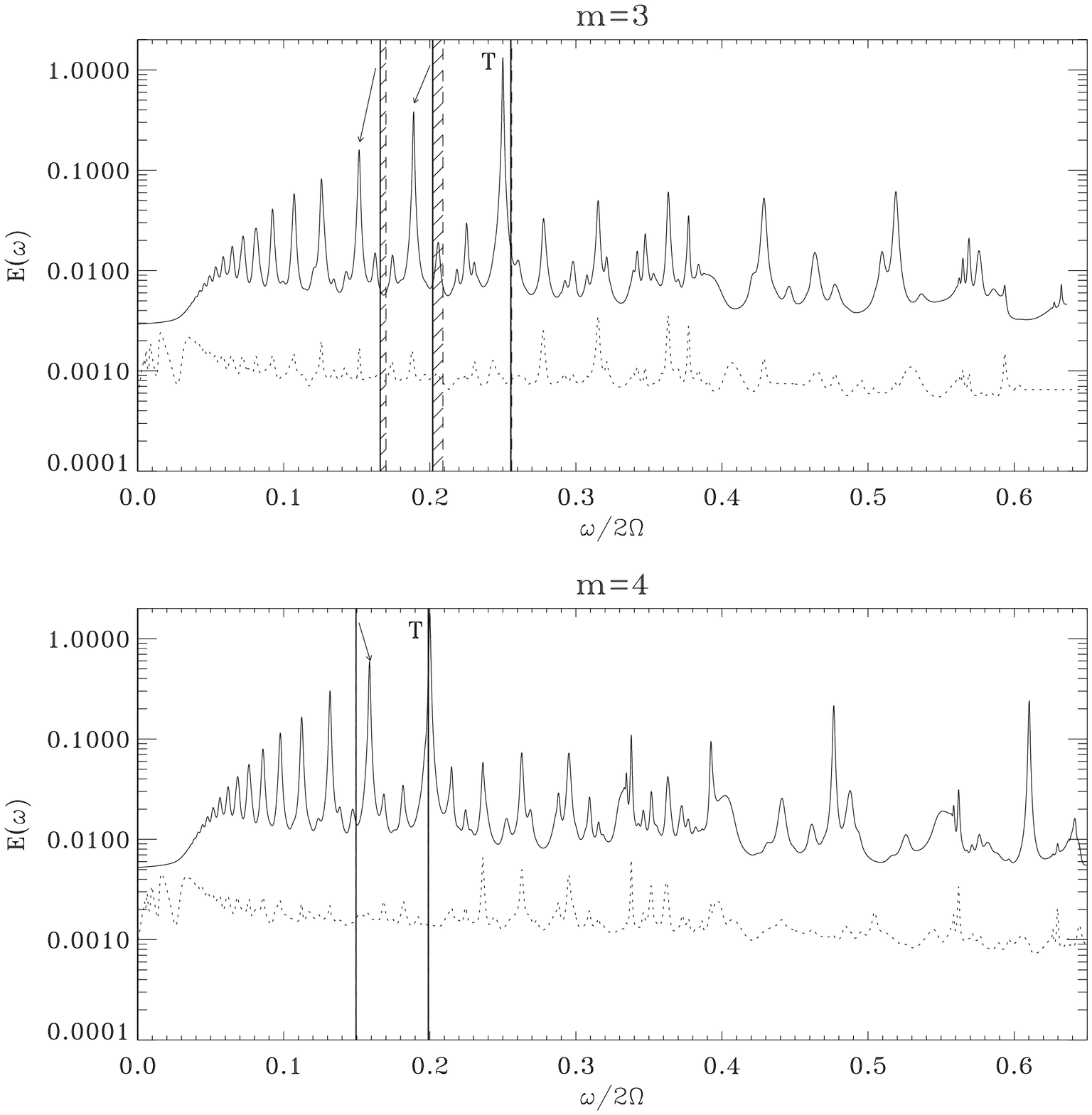}}
\caption[]{Same as Fig.~\ref{couette_flow1} but for $m=3$ and $m=4$.}
\label{couette_flow2}
\end{figure}

We solve these equations numerically, following the same spectral method as in
\cite{RV10}. The fields are first expanded into spherical harmonics as
follows:

\[\vu=\sum_{l=0}^{+\infty}\sum_{m=-l}^{+l}\ulm(r)\RL+\vlm(r)\SL+\wlm(r)\TL
,\]
with

\[\RL=\YL(\theta,\varphi)\vec{e}_{r},\qquad \SL=\na\YL,\qquad
\TL=\na\times\RL \]
where gradients are taken on the unit sphere. The radial $\ulm$ and $\wlm$
function thus verify

\greq
E\Deltal\wlm - i\hat{\omega} \wlm= \\
 \hspace{0.1cm} -A_{\ell}r^{\ell-1}\dr{} \biggl(
\frac{\ulmm}{r^{\ell-2}}\biggr)
-A_{\ell+1}r^{-\ell-2}\dr{}\biggl( r^{\ell+3}\ulp\biggr) \\
\\
E\Deltal\Deltal(r\ulm)-i\hat{\omega} \Deltal(r\ulm)= \\
\hspace{0.1cm} B_{\ell}r^{\ell-1}\frac{\partial}{\partial r}
\biggl(\frac{\wlmm}{r^{\ell-1}}\biggr) + B_{\ell+1}r^{-\ell-2}
\dr{}\biggl( r^{\ell+2}w^{\ell+1}_m\biggr)
\egreqn{eqproj}
for $\ell\in[m,L]$. $L$ is the order of the truncation in the spherical
harmonics expansion. $\vlm$ is eliminated using mass-conservation, and
we introduced

\[ A_\ell = \frac{1}{\ell^2}\sqrt{\frac{l^2-m^2}{4\ell^2-1}}, \qquad
B_\ell = \ell^2(\ell^2-1)A_\ell,\]

\[ \Deltal = \frac{1}{r}\ddnr{}r - \frac{\ell(\ell+1)}{r^2} \]
The radial functions are then discretized on Gauss-Lobatto collocations
nodes. Equations are completed by boundary conditions \eq{bc1} and
\eq{bc2} leading to a linear system like $[A]x=b$, which is solved by
classical numerical methods. Note that these numerical solutions do not
take into account the shaft bearing the inner core.

\begin{figure}
\centerline{
\includegraphics[width=\linewidth,clip=true]{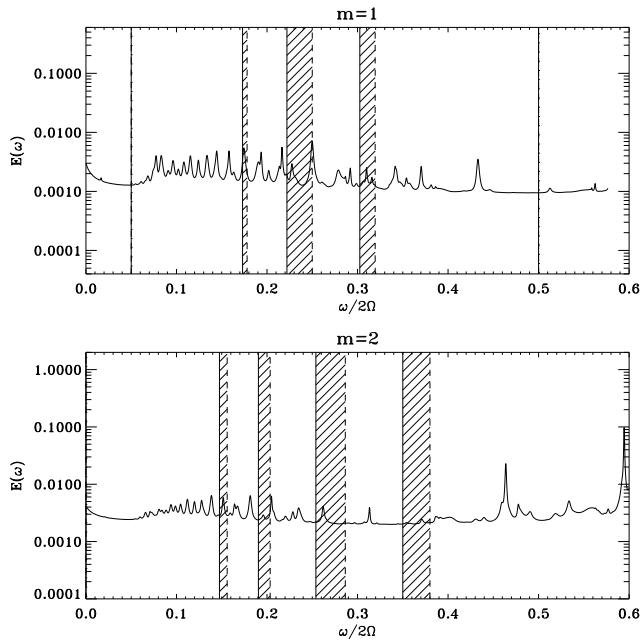}}
\caption[]{Same as Fig.~\ref{couette_flow1} but for equatorially symmetric modes with
$m=1$ and $m=2$.}
\label{couette_flow1+}
\end{figure}

\begin{figure}
\centerline{
\includegraphics[width=\linewidth,clip=true]{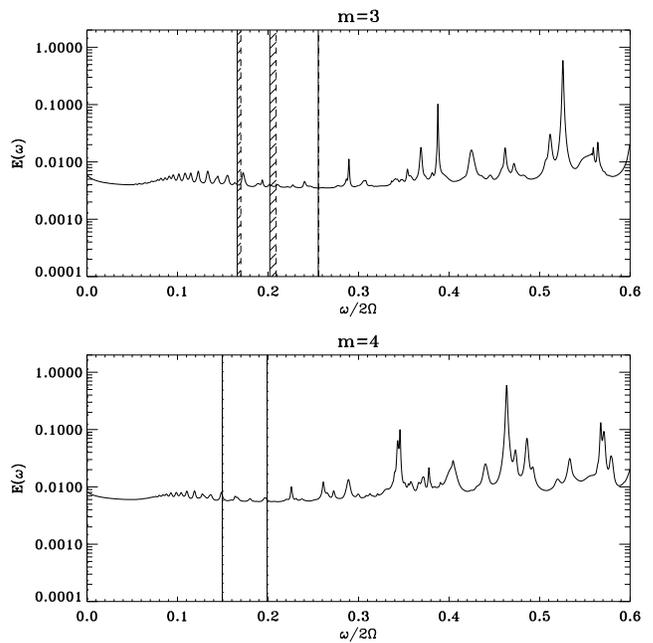}}
\caption[]{Same as Fig.~\ref{couette_flow1+} but for $m=3$ and $m=4$.}
\label{couette_flow2+}
\end{figure}

\subsection{Results from the model}

The above model is very simple since it just reproduces the geometry of
the container, the background rotation outside the tangent cylinder and
the effects of viscosity. No differential rotation is included. Thus,
from the comparison of its results with the data we only expect an
identification of the modes and a comparison of the width of the
resonances (i.e. of the dissipative processes).

Solving \eq{eqmotion} with \eq{bc1} and \eq{bc2}, we
computed the resonance curves for various azimuthal symmetries,
namely $m=1,2,3,4$. Fig.~\ref{couette_flow1} and \ref{couette_flow2}
show the total kinetic energy of the responding oscillations in
some arbitrary unit as a function of the forcing frequency. We show
the response of the fluid when the excitation is imposed either
on the inner boundary or the outer boundary.  Obviously, the outer
boundary forcing is much more efficient than the inner boundary one.
The hatched bands correspond to the observed range of frequencies
of the oscillations when the Rossby number is varied. As shown by
the spectrograms (Fig.~\ref{spectrogram1}-\ref{spectrogram2} and
Fig.~\ref{omega_ro1}-\ref{omega_ro2}), the frequency of the modes
systematically decreases when $|\RO|$ decreases, that is when the
background flow gets closer to the solid body rotation. Therefore,
only the left boundary of the frequency bands should be compared to the
frequencies of the model. Thus doing, we can identify clearly the purely
toroidal modes for $m=1,2,3,4$ whose frequencies are

\begin{figure}
\centerline{
\includegraphics[width=0.50\linewidth,clip=true]{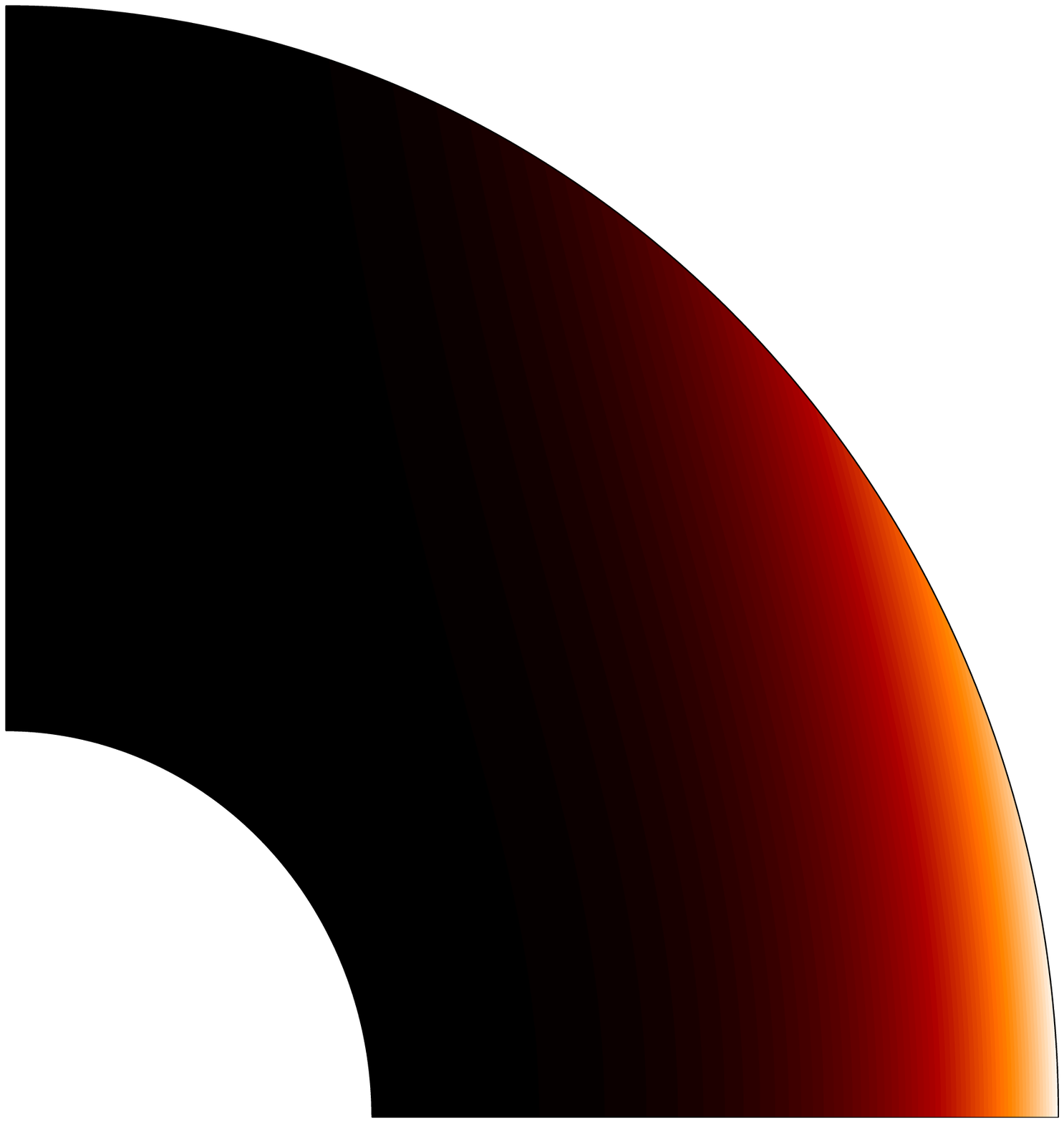}
\includegraphics[width=0.50\linewidth,clip=true]{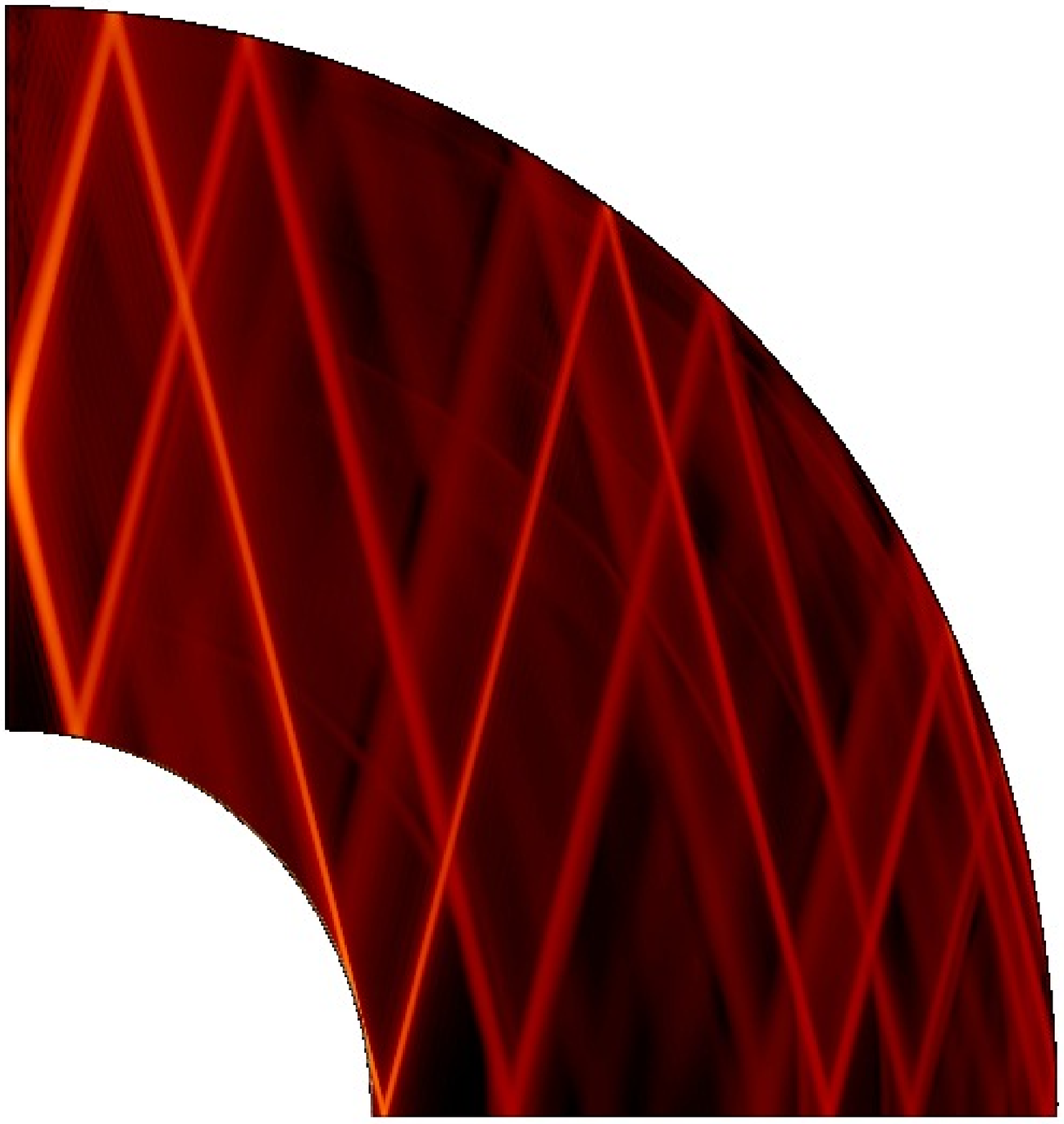} }
\caption[]{(Color online) Left: the kinetic energy distribution in a meridional
plane of the flow associated with the resonance of the $m=3$ mode at
$\hat{\omega}=0.25$. The Ekman number used for this calculation is 2.5$\times10^{-9}$. The
numerical resolution is $L=780$ (highest spherical harmonic order) and $N_r=300$ (highest
Chebyshev polynomial order). Right: the same as left but when the large-scale
part of the flow field described by \eq{ls} is filtered out.}
\label{dissec}
\end{figure}

\[ \hat{\omega}=\frac{\omega}{2\Omega} = \frac{1}{m+1}\]
For these frequencies, an analytical solution exists \cite[][]{RV97} and reads

\greq
v_\theta = A r^m (\sin\theta)^{m-1}\sin(m\phi+\omega_m t) \\
v_\phi = A r^m (\sin\theta)^{m-1}\cos\theta\cos(m\phi+\omega_m t)
\egreqn{ls}
if viscosity is neglected. When viscosity is included the structure of
these modes is more complex since shear layers are excited by the critical
latitude singularity. In Fig.~\ref{dissec} we display the kinetic
energy of the flow at $\omega/2\Omega=1/4$ (m=3) and when the large-scale part
expressed in \eq{ls} is filtered out. One clearly sees the dissipative
structures which are associated with the viscous stress.

Back to Fig.~\ref{couette_flow1} and \ref{couette_flow2}, we note that the
frequency mismatch between model and experiment is less and less as $m$
increases for the purely toroidal modes. This is because modes of high
$m$ are more concentrated in the equatorial region of the outer shell
(as may be seen from Eq.~\ref{ls}) and therefore less affected by the
differential rotation.

Other modes with $m=2$ or $m=3$, may also be identified with the resonance
peaks which stay on the left of the band. Indeed, the resonances
of the model are valid for the asymptotic case $\RO\tv0$, while the
observed frequencies decrease with $|\RO|$. For other non-toroidal modes
(with m=1 and m=4) identification is only tentative.

Besides, the model gives the width of the resonances. Table~\ref{fwhm} shows
the full width at half-maximum (FWHM) of the identified modes and
the corresponding value of the model. We note that for half of the
observed modes, the comparison is good. For the remaining modes, either
the peaks are much wider or much narrower than those of the model. In
a linear model, the width of the peaks is related to the dissipative
process at work or to a casual blend of another neighbouring peak. Wider
experimental peaks means a higher dissipation in the experiment than in
the model, which is understandable: the experimental flow may contain
small-scale turbulence that can increase dissipation.  The opposite is
more surprising. It may mean either a misidentification or a real change
of the important scales of the modes when the differential rotation
is strong.  We actually note that the mismatch is larger for the m=1
resonances, which exist at very negative Ro.

\begin{figure}
\begin{center}
\includegraphics[width=0.98\linewidth,clip=true]{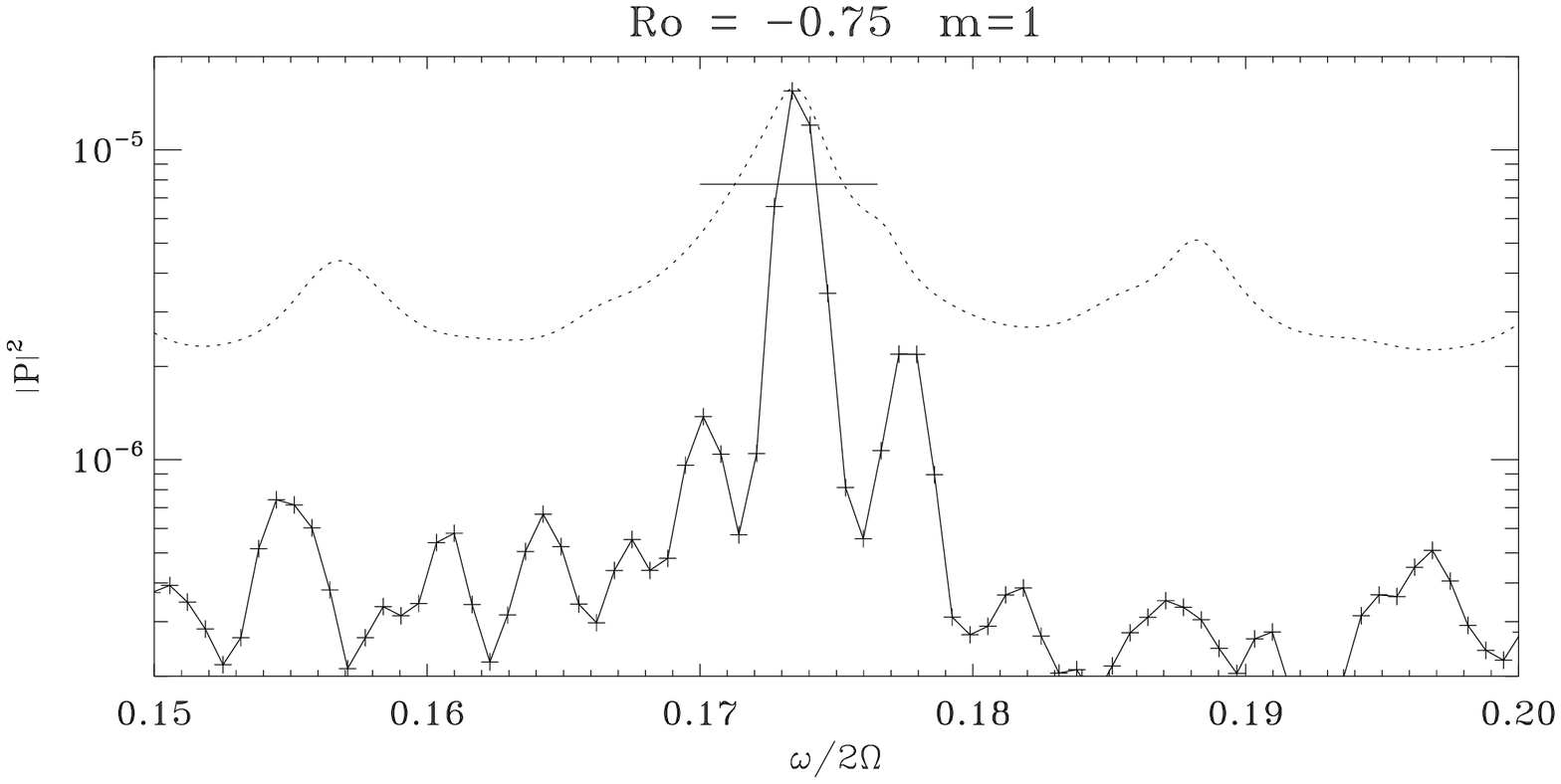}  \\
\includegraphics[width=0.98\linewidth,clip=true]{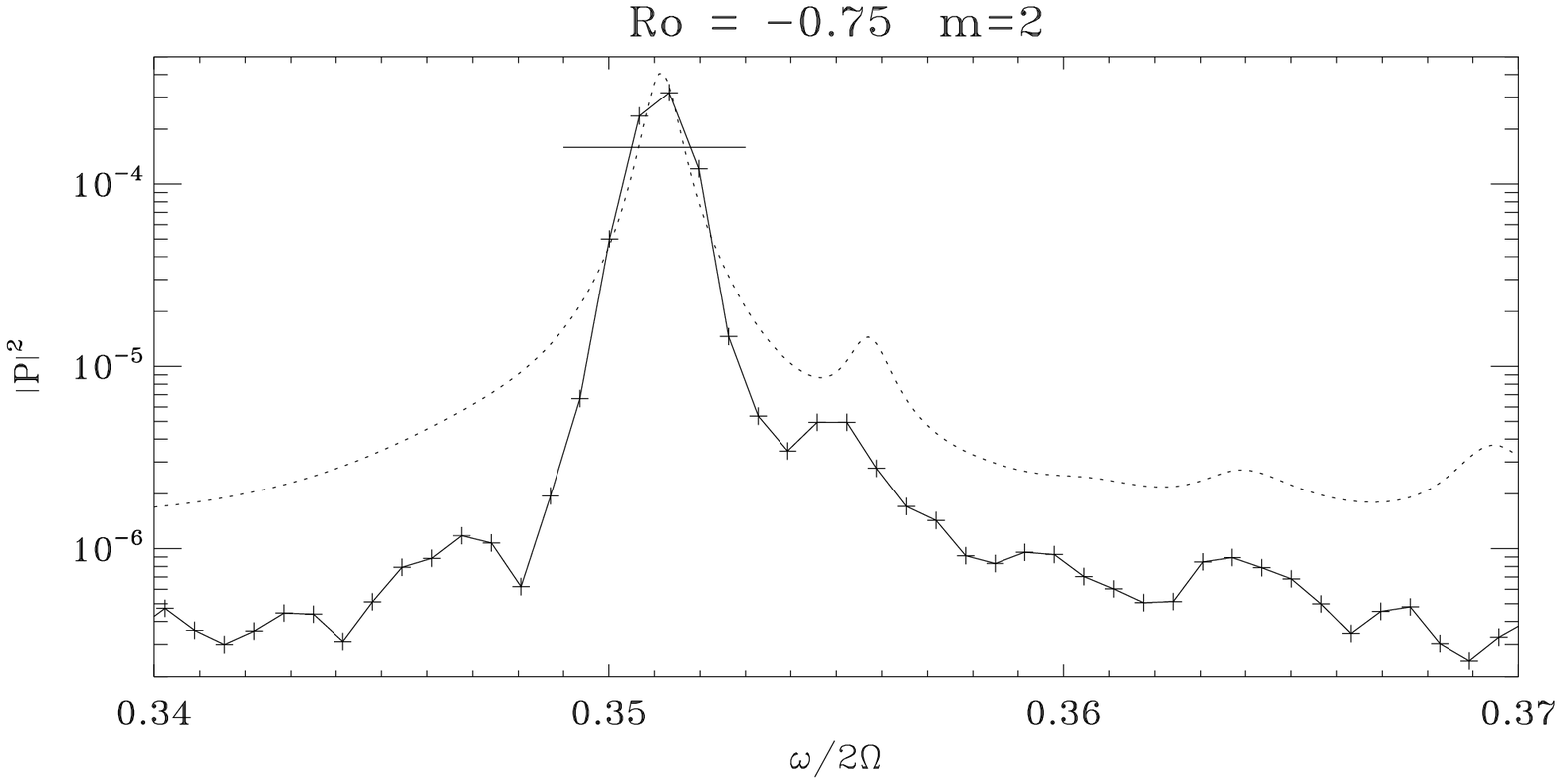} 
\end{center}
\caption[]{Comparison of the width of two observed resonances
at Ro=-0.75 (solid lines with pluses) with the prediction of the
model (dotted lines). The model curve has been shifted in frequency so
as to match the observed frequency. The top figure is centered on the m=1
mode at $\hat{\omega}\sim0.17$, while the bottom one shows the resonance
of the purely toroidal m=2 mode. The horizontal line gives the position
of the half-maximum. $E=2.5 10^{-8}$.}

\label{figfwhm}
\end{figure}

\begin{table}
\begin{center}
\begin{tabular}{|cccccc|}
\hline
$m$  &  $\hat{\omega}$ & $\delta\hat{\omega}\times10^{4}$ & Ro &
$\delta\hat{\omega}_{\rm model}\times10^{4}$ &
$\delta\hat{\omega}/\delta\hat{\omega}_{\rm model}$ \\
 &&&&& \\
1  & 0.31 & 29 & -1.75 & 12 & 2.4 \\
1  & 0.23 & 8 & -1.13 & 56 & 0.14 \\
1  & 0.17 & 9 & -0.80 & 38 & 0.24 \\
 &&&&& \\
2  & 0.3526 & 8 & -0.80 & 8 & 1.0 \\
2  & 0.2628 & 12 & -0.60 & 10 & 1.2 \\
2  & 0.1980 & 23 & -0.43 & 29 & 1.26 \\
2  & 0.1667 & 20 & -0.33 & 22 & 0.93 \\
 &&&&& \\
3  & 0.2573 & 20 & -0.4 & 10 & 2.0 \\
3  & 0.2035 & 12 & -0.3 & 12 & 1.0 \\
3  & 0.1662 & 9 & -0.23 & 16 & 0.55\\
\hline
\end{tabular}
\end{center}
\caption[]{Comparison of the experimental FWHM ($\delta\hat{\omega}$)
of the mode at frequency $\hat{\omega}$ of azimuthal wavenumber $m$,
measured at Rossby number Ro, with the FWHM predicted by the model. The
last column gives the ratios of the experimental and numerical FWHMs.}
\label{fwhm}
\end{table}

In Fig.~\ref{figfwhm}, we show two resonances which appear at $\RO=-0.75$. One
is associated with m=1. The model curve has been shifted so that peaks
frequencies coincide. We see that the widths poorly match.  The other
mode visible at this Rossby number, is the m=2-purely toroidal one.
Its observed frequency is $\hat{\omega}\sim 0.35$, expected at
$\hat{\omega}=1/3$ for a solid body rotation. Here too we shift the
model curve and note that the widths of the peaks better match. Moreover, we
also notice a tiny peak on the high frequency side of the main resonance
that appears both in the experimental data and in the model, suggesting
that we have correctly identified this mode.

To further secure mode identification, we also investigated the
other possible symmetry of the modes.  Indeed, in equatorially
symmetric containers, two independent set of modes coexist for
a given azimuthal wavenumber $m$: modes for which the pressure
function is symmetric with respect to equator and those for which
it is anti-symmetric. For these latter modes the pressure and the
azimuthal velocity component $v_\varphi$ vanish at equator. We thus
computed the analog of Fig.~\ref{couette_flow1},\ref{couette_flow2}
but for equatorially symmetric modes. Results are shown in
Fig.~\ref{couette_flow1+},\ref{couette_flow2+}. Quite clearly, none
of the m=3 or m=4 modes can be related to experimental resonances. For
the m=1 and m=2 azimuthal wavenumbers, the two lowest frequencies may
possibly correspond to symmetric modes.

To conclude on mode identifications, we may say that the vicinity of
frequencies together with the matching of the width of the resonances
allow us to identify almost all the observed frequencies with the
eigenfrequencies of equatorially antisymmetric retrograde inertial
modes of spherical shell in solid rotation. Doubts may concern two
modes actually: m=1, $\hat{\omega}=0.17$ and $\hat{\omega}=0.23$,
because they show experimental FWHMs which are significantly less than
those predicted by the model. For these two modes the background shear
may have more important consequences, changing for instance the length
scales  that control viscous dissipation (in addition, two equatorially
symmetric modes have similar frequencies, thus possibly adding some
confusion). Hence, we shall retain that in this experimental spherical
Couette flow, excited inertial waves are generally (if not systematically)
equatorially antisymmetric.

\begin{figure}
\centerline{
\includegraphics[width=0.98\linewidth,clip=true]{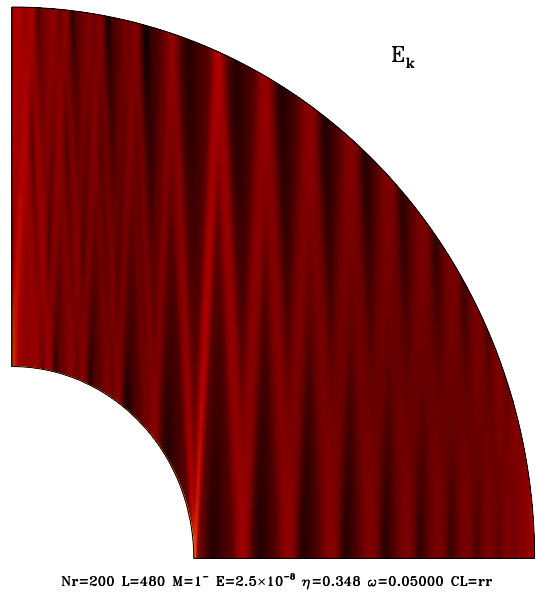} }
\caption[]{(Color online) The kinetic energy distribution in a meridional
plane of the flow associated with the resonance at $\hat{\omega}=0.05$
for m=1.}
\label{dissec_sst}
\end{figure}

Finally, let us note that the spectrograms show that excited inertial
modes live in a definite interval of Rossby numbers. We may even observe
that for a given $m$, when the Rossby number is decreased, modes appear
one by one, the ``old" one disappearing when a new one appears. The
series ends when the purely toroidal one associated with the given $m$
(see Eq.~\ref{ls}) is excited. We note however that the spin-over mode
at $\hat{\omega}=0.5$ and a (or a set of) low-frequency resonance(s)
with $m=1$ around $\hat{\omega}=0.05$ do not follow this rule, being
permanently excited when $\RO>-1$.

The resonance near $\hat{\omega}=0.05$ may be associated with the broad
peak appearing on the m=1-curve at $\hat{\omega}\sim0.03$, when the
forcing is set on the inner core.  We show in Fig.~\ref{dissec_sst}
the shape of the oscillating flow at this frequency for a solid body
rotation. It looks like a set of shear layers spawned by the critical
latitude singularity on the inner boundary.

\section{Discussion}

Let us first summarize the foregoing conclusions derived from the
combination of experimental and numerical results. The spherical Couette flow
with $E=2.5\, 10^{-8}$ and $-1.8 < \RO=\frac{\Omega_i}{\Omega}-1 <0$
(the inner sphere rotating more slowly than the outer sphere) displays a
series of excited inertial modes which have the following properties:

\begin{enumerate}
\item They are all non-axisymmetric, with an azimuthal wavenumber
$m\in\{1,2,3,4\}$.
\item They are likely all anti-symmetric with respect to equator.
\item They all propagate azimuthally in the opposite direction of the
outer shell rotation (\ie they are retrograde).
\item Their highest observed frequency is that of the purely toroidal
mode (see Eq.~\ref{ls}) associated with the given azimuthal wavenumber $m$,
namely $\omega=2\Omega/(m+1)$.

\item When the differential rotation is increased from zero (\ie when
decreasing the Rossby number from zero), inertial modes of a given $m$
turn on at a specific Rossby number and turn off when the next mode of the
series turns on. The spin-over mode is an exception, being always
excited when $\RO>-1$.
\end{enumerate}

We understand the emergence of these inertial modes in the experimental
Couette flow as a consequence of the existing shear layer staying on the
tangential cylinder, namely the layer which replaces the Stewartson
layer when $\RO$ is not small. Our scenario is as follows.

\begin{figure}
\centerline{
\includegraphics[width=\linewidth,clip=true]{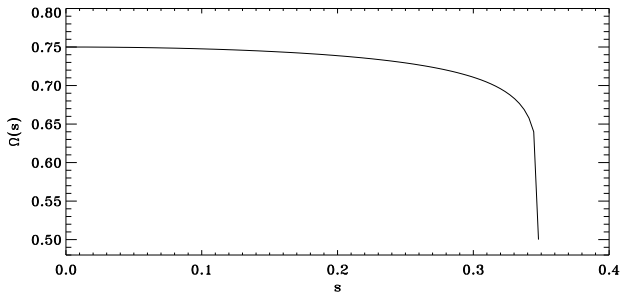}}
\caption[]{View of the differential rotation inside the tangential
cylinder as given by the linear solution of \cite{Prou56}. Here
$\Omega_i=0.5$, $\Omega=1$ and $\eta=0.348$. Note that the angular
velocity decrease is a feature of the inviscid solution that is
compensated by the rapid rise in the Stewartson layer
\cite[see][for a complete view with viscous effect]{DCJ98}.}
\label{difrot}
\end{figure}

Let us first recall that at very small Rossby numbers (in the linear
regime), and in the limit of vanishing Ekman numbers, the spherical
Couette flow may be represented by a solid body rotation outside the
tangent cylinder and a differential rotation inside the tangent
cylinder. Weak circulations and the Stewartson layer complete these
basic azimuthal flows \cite[][]{DCJ98}. Proudman\cite{Prou56} has shown that
inside the tangent cylinder

\beq \Omega(s) = \Omega_i+(\Omega-\Omega_i)
\frac{(1-s^2/\eta^2)^{1/4}}{(1-s^2/\eta^2)^{1/4} +(1-s^2)^{1/4}}
\eeqn{rotdiff}
where $s$ is the radial cylindrical coordinate. The main point shown by
this formula is that the fluid inside the tangent cylinder rotates
almost rigidly at an intermediate angular velocity:

\[ \Omega(0) = \frac{\Omega_i+\Omega}{2}\]
Fig.~\ref{difrot} further illustrates this point showing that the
angular velocity does not vary very much except near the Stewartson layer.

Eq.~\eq{rotdiff} has been derived in the asymptotic limit of small numbers
and it is certainly approximate when applied to the experiment. However,
it seems reasonable to represent the actual flow by a solid body rotation
at angular velocity $\Omega$ outside the tangent cylinder and at angular
velocity $(\Omega_i+\Omega)/2$ inside it. A shear layer, generalizing
the Stewartson one connects these two rotations.

\begin{figure}
\centerline{
\includegraphics[width=0.98\linewidth,clip=true]{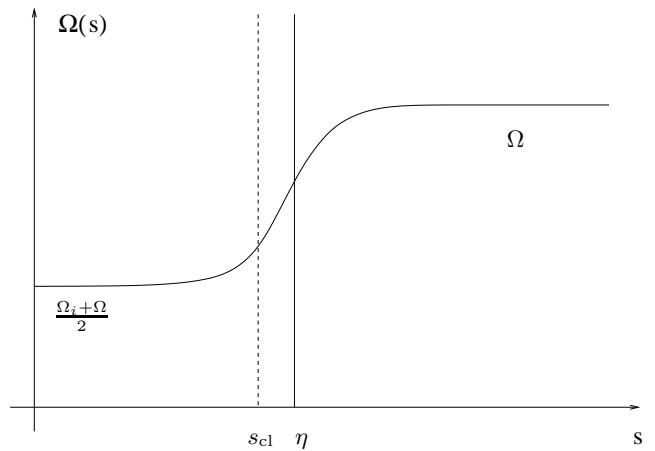}
}
\caption[]{Schematic view of the bulk rotation of the fluid in
the spherical Couette flow for $\eta < z < \sqrt{1-\eta^2}$
($0.348<z<0.937$). The transition between the outer solid body rotation
and the inner (approximate) solid body rotation is along the tangent
cylinder at $s=\eta$. When $\Omega_i$ is decreased, the critical
layer of a given mode first appears on the left side of the shear layer,
and moves towards larger $s$.}
\label{rotprof}
\end{figure}

Using the $(2\Omega)^{-1}$ time scale, the angular velocity of fluid
$\Omega(s)$ thus varies in the interval $[\RO/4+1/2, 1/2]$. Hence, in
the foregoing experiment where $\RO<0$, the flow is retrograde inside
the tangent cylinder as viewed from the outer shell. This implies that
inertial waves propagating outside the tangent cylinder and with a
retrograde angular velocity, face a critical layer if their angular
phase velocity $-\hat{\omega}/m$ is less (in absolute value) than
$\RO/4$. In other words, if

\beq \RO < -\frac{4\hat{\omega}}{m}<0\eeqn{ccl}
is verified, then there exist a place within the shear layer where the phase
velocity of the wave equals that of the fluid. This is the place where
a critical layer develops. In Fig.~\ref{rotprof} we sketch out the position of the
critical layer. Such a layer may over reflect the inertial waves
and be at the origin of the selection of excited modes as proposed by
Kelley et al. \cite{KTZL10}. Critical layers are indeed known to play a crucial
part in the dynamics of the Rossby waves in the Earth atmosphere
\cite[][]{haynes03,KM85}. Recently, Baruteau and Rieutord \cite[][]{BR12}
found that some inertial modes propagating over a differentially rotating
fluid in a spherical shell can be unstable when a critical layer exists. We
shall now precise this scenario.

\psfrag{omm}{$\frac{\Omega_i+\Omega}{2}$}
\psfrag{om}{$\Omega$}
\psfrag{scl}{$s_{\rm cl}$}
\psfrag{eta}{$\eta$}

\begin{figure}
\centerline{
\includegraphics[width=1.00\linewidth,clip=true]{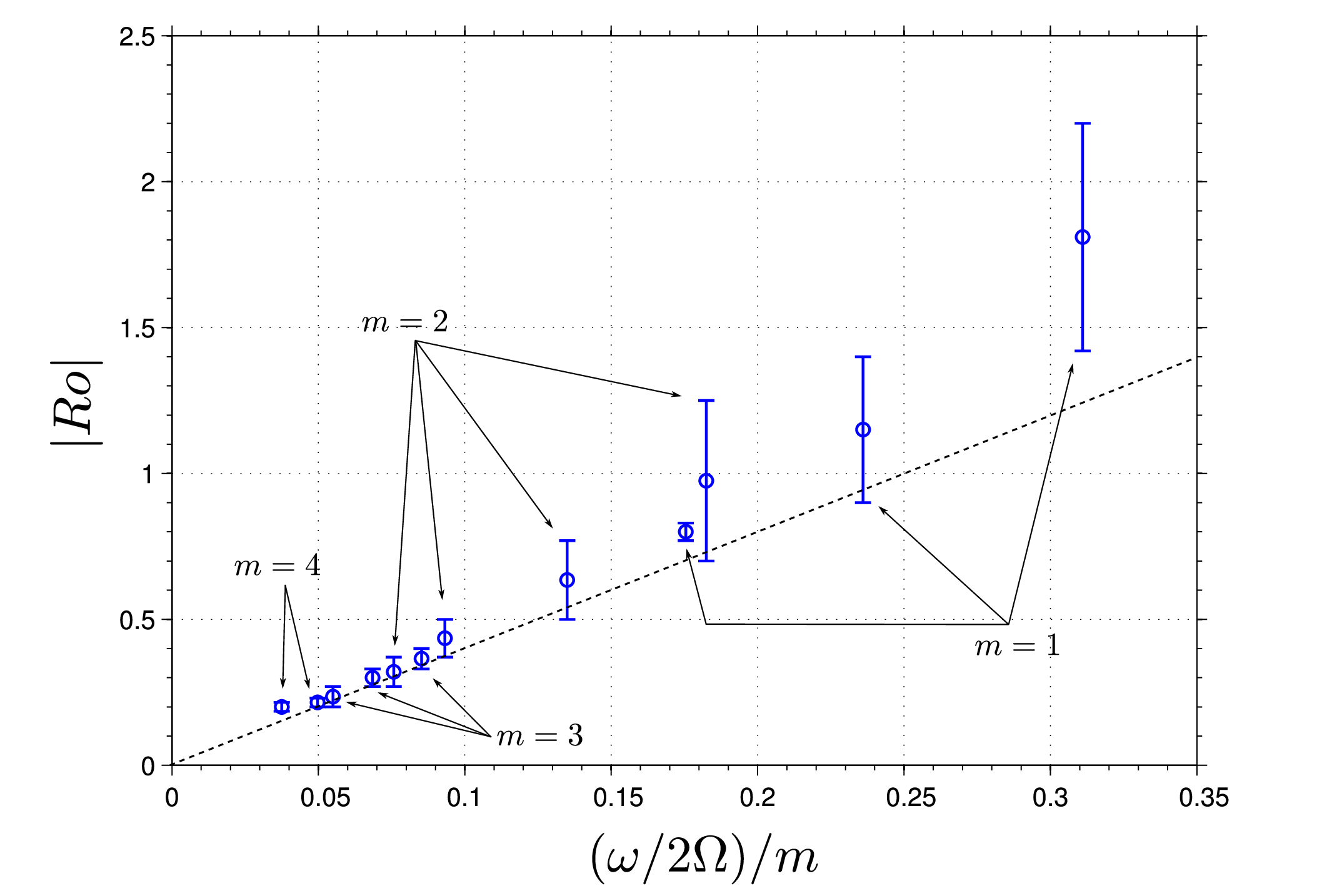}
}
\caption[]{(Color online) Graphic illustration of Tab.~\ref{rocrit}:
for each observed inertial mode, we display the Rossby number range
of existence of the mode. The dashed line shows the (absolute value)
of the critical Rossby number $4\hat{\omega}/m$ below which the mode is
no longer excited. Note the good matching with the experimental value.}
\label{romega}
\end{figure}

Eq.~\eq{ccl} shows that for a given (retrograde) non-axisymmetric
inertial modes, there is a critical negative Rossby number below which
the critical layer exists and might excite the mode. This critical
Rossby number can be derived from the resonances frequencies showing up in
Fig.~\ref{couette_flow1} and \ref{couette_flow2}, and compared to the
maximum Rossby number beyond which the mode is no longer excited. This
Rossby number is derived from the spectrograms in Fig.~\ref{spectrogram1}
and \ref{spectrogram2}. The results of this exercise is shown in
Tab.~\ref{rocrit} and illustrated with Fig.~\ref{romega}.  The matching
is remarkable, especially for the m=2 and m=3 modes.

\begin{table}
\begin{center}
\begin{tabular}{|ccc|c|c|}
\hline
$\quad m$\quad  & $\quad\hat{\omega}\quad$ & $\quad-4\hat{\omega}/m\quad$ &
$\quad\RO_{\rm crit}\quad$ & $\alpha$ \\
\hline
&&&& \\
1   & 0.1595         & -0.638 & -0.74&  \\
1   & 0.210          & -0.840 & $>-1.14$& 0.27 \\
1   & 0.301          & -1.204 & -1.39   & 0.11\\
&&&& \\
\hline
&&&& \\
2   & 0.142	     & -0.284 & -0.29 &      \\
2   & 0.177	     & -0.355 & -0.36 & 0.275\\
2   & 0.233	     & -0.466 & -0.50 & 0.236\\
2   & 0.333	     & -0.667 & -0.70 & 0.107\\
&&&& \\
\hline
&&&&\\
3   & 0.152	     & -0.203 & -0.20 & \\
3   & 0.189	     & -0.252 & -0.25 & \\
3   & 0.250	     & -0.333 & -0.33 & \\
&&&& \\
\hline
\end{tabular}
\end{center}
\caption[]{Table of the critical Rossby numbers ($\RO_{\rm crit}$)
as derived from the experimental spectrograms compared to the
predictions of the numerical model $-4\hat{\omega}/m$. $\alpha$ is a
parameter controlling the frequency drift with $\RO$ and derived from
Figs.~\ref{omega_ro1},\ref{omega_ro2}.}
\label{rocrit}
\end{table}

When the Rossby number is decreased from zero, large-scale resonances
appear in the order given by Fig.~\ref{couette_flow1} and
\ref{couette_flow2}, until the purely toroidal mode of the associated
$m$ is excited. The m=1-purely toroidal spin-over mode is an exception
as it consists of a global solid body rotation around an equatorial
axis, oscillating at the rotation frequency of the outer
shell. Therefore, it can be mechanically excited by the boundary.

The spectrograms also show that when a new mode is excited the
``old" one with the same $m$ disappears. We understand this process
as follows: the new mode is always of larger scale than the old one
(Fig.~\ref{couette_flow1} and \ref{couette_flow2} show that the strength
of the resonances increases as the frequency increases), thus when the
Rossby number decreases a less-damped mode takes over a more damped
one. In the differentially rotating fluid where these modes are unstable
because of the critical layer, a more unstable mode overwhelms a less
unstable one. This behaviour is typical of instabilities where many
modes are destabilized simultaneously and the most unstable dominates the
bifurcated state (e.g. Rayleigh-B\'enard instability or Taylor-Couette
instability). The change of excited mode stops when the least-damped of
the series, the purely toroidal one, is excited.

This scenario thus explains why only non-axisymmetric retrograde modes
are excited, and the Rossby number below which this is indeed the
case. However, it does not explain why only anti-symmetric modes
are excited. We conjecture that the coupling with the spin-over mode
may be the key of this selection. Indeed, from the work of Hollerbach
\cite{H03} we know that the Stewartson layer generated by the spherical
Couette flow can become unstable when the Rossby number is increased
(in absolute value). Modes with the same symmetry as the original
flow, namely symmetric with respect to equator, are destabilized.
Their wavelength should scale with the width of the layer, namely
$E^{1/4}$. From the numerical results of \cite{H03}, we infer that the
most unstable mode has an azimuthal wavenumber $m\sim0.465E^{-1/4}$. In
the context of the experiment described above, such modes have $m\sim16$
(E=7.5$\times10^{-7}$) or $m\sim37$ (E=2.5$\times10^{-8}$).  Hence, the
linear instability injects energy at rather high wavenumbers implying,
thanks to nonlinear interactions and an inverse cascade of energy, that a
large range of scales (with azimuthal wavenumbers $0 < m \infapp 40$)
is excited, thus generating some turbulence. The equatorial symmetry
could be broken by the spin-over mode (namely the $m=1$-mode described by
Eq.~\ref{ls}), whose presence has been detected on the velocity signal.
This mode is anti-symmetric.  The nonlinear interaction of this  mode
with disturbances growing over the turbulent Stewartson layer could thus
select the observed antisymmetric inertial modes. However, more detailed
numerical investigations are definitely necessary to answer this question.

Finally, in light of the foregoing scenario, we can also interpret
the linear drift of the frequencies with the Rossby number as shown
in Fig.~\ref{omega_ro1} and \ref{omega_ro2}. The linear dependence
suggests that

\beq \hat{\omega} = \hat{\omega}_{\rm crit} - \alpha\frac{m}{4}\lp \RO-\RO_{\rm
crit}\rp\eeqn{omro1}
where $\RO_{\rm crit}=-4\hat{\omega}_{\rm crit}/m$ and $\alpha$ is a
constant. With the expression of $\RO_{\rm crit}$ we may rewrite the
previous equation as

\beq \hat{\omega} = \hat{\omega}_{\rm crit}(1-\alpha) -\frac{\alpha m}{4}\RO
\eeqn{omro}
which shows that $\alpha$ should adjust both the slope and the frequency
at $\RO=0$.  Fitting experimental data, we derive the value of $\alpha$
for the modes used in Fig.~\ref{omega_ro1} and \ref{omega_ro2}. These
are given in Tab.~\ref{rocrit}. The physical meaning of $\alpha$ may
be derived from our scenario. Indeed, if the modes are excited through
their critical layer, their azimuthal phase velocity $-\hat{\omega}/m$
equals the fluid angular velocity somewhere in the shear layer. We
recall that the scaled angular velocity of the fluid viewed from
the frame co-rotating with the outer shell varies from 0 down to
$\RO/4<0$. Introducing the angular velocity profile $\tilde{\Omega}(s)$
such that $1\geq\tilde{\Omega}(s)\geq0$, we have

\[\hat{\Omega}(s,\RO) = \frac{\RO}{4}\tilde{\Omega}(s) .\]
In our modelling of the spherical Couette flow, we assume that the
rotation profile is independent of $\RO$.

If $s_{\rm cl}$ is the position of the critical layer, then
$-\hat{\omega}/m=\frac{\RO}{4}\tilde{\Omega}(s_{\rm cl})$. Using \eq{omro}
leads to

\beq \tilde{\Omega}(s_{\rm cl}) = \alpha + (1-\alpha)\frac{\RO_{\rm crit}}{\RO}\eeq
This expression shows that the constant $\alpha$ gives the
asymptotic position of the critical layer when $\RO\tv-\infty$. Since
$\tilde{\Omega}(s)$ is a decreasing function of $s$, it also shows
that, as the differential rotation increases ($\RO$ more negative),
the critical layer moves away from the rotation axis ($s_{\rm cl}$
increases). Furthermore, modes with a higher $\alpha$ have critical
layers more inwards the tangent cylinder.

In Sect.~\ref{exp}, we noted that the first m=1-mode shows a
different dependence with \RO\ when the Ekman number $E$ is doubled
(approximately). From \eq{omro1}, this translate into an increased
$\alpha$ for a lowered $E$. Unfortunately, there is no univocal
interpretation of this change: it may either mean that the critical layer
moved inside the shear layer or that the rotation profile changed, or
both. Either a more elaborated model or more precise measurements using
other modes is necessary to disentangle the effects.

\section{Conclusions}

In this work we have been able to refine the scenario proposed by Kelley
et al. for the selection of inertial modes in a spherical Couette flow
\cite[][]{KTZL10}. We showed that the observed modes are most likely
excited by a critical layer lying inside the shear layer which separates
the fluid inside and outside the tangent cylinder. We demonstrated that
this mechanism leads to a negative critical Rossby number, below which
some non-axisymmetric retrograde inertial modes can be excited. The
predicted value of this critical Rossby number matches quite nicely
the observed experimental values, thus giving support to the proposed
mechanism. However, because of the strong differential rotation needed to
excite some modes, the identification of a few oscillation frequencies
with those of a fluid inside a  spherical shell in solid body rotation
remains uncertain. Another pending question is that of the equatorial
symmetry of the observed modes. Our scenario along with the matching of
frequencies of many modes argue in favour of a selection of anti-symmetric
modes. However, this conclusion is not completely firm because some of
the symmetric modes may match a few of the frequencies and also because
the FWHM of two resonances show experimental values that
are much less than those predicted by the model, which is {\it a priori}
less dissipative.

These results underline the need of more detailed investigations of
inertial oscillations within differentially rotating fluids.  Preliminary
numerical results \cite[][]{BR12} show that a global shear can indeed
significantly modify the properties of inertial modes and also shows that
critical layers can destabilize some non-axisymmetric modes. However,
the very case of the mean flow associated with a quasi-turbulent spherical
Couette flow remains to be investigated.

Such a study may lead to the interesting perspective of reconstructing
the interior differential rotation of the fluid by adjusting the
prediction of the model to the observed value of the resonance
frequencies. One could thus deduce details of the mean flow. Such a
technique is similar to helioseismology techniques in solar research
\cite{thompson_etal96}, but instead using inertial modes to reconstruct
the internal rotation of the fluid.  Inertial modes are certainly {the
most appropriate} modes to infer the rotational property of a star or
a planet. Our comparison of the inertial frequencies of our model with
the observed ones, shows that a laboratory spherical Couette flow offers
a unique playground to test this use of inertial waves.

Finally, let us point out that the ultimate challenge of modelling (this) experimental
spherical Couette flows is to predict the amplitude of excited modes as a function of
the Rossby number and to reproduce the sequence of their appearance.

\begin{acknowledgements}
The numerical calculations have been carried out on
the NEC SX8 of the `Institut du D\'eveloppement et des Ressources en
Informatique Scientifique' (project 020666) and on the CalMip machine of the
`Centre Interuniversitaire de Calcul de Toulouse' (project 0107), which are both
gratefully acknowledged. The University of Maryland team acknowledges
support from the U.S. National Science Foundation EAR 1114303.

\end{acknowledgements}

\bibliographystyle{apsrev4-1}
\bibliography{../../../biblio/bibnew}

\end{document}